\documentclass[twocolumn,showpacs,showkeys,reprint,amsmath,amssymb,aps,prm,longbibliography]{revtex4-2}
\usepackage{graphicx}% Include figure files
\usepackage{dcolumn}% Align table columns on decimal point
\usepackage{bm}% bold math
\usepackage{tabularx}% Table width
\usepackage{amsmath}% Higher math
\usepackage{xcolor}
\usepackage{hyperref}
\usepackage{newtxtext} % 
\usepackage{newtxmath} % 
\hypersetup{
    colorlinks=true,
    linkcolor=blue,
    filecolor=blue,
    urlcolor=blue,
    citecolor=blue
}

\emergencystretch=\maxdimen
\begin{document}

\title{Pressure-induced double-dome superconductivity in doped kagome metal Cs(V$_{0.86}$Ta$_{0.14}$)$_3$Sb$_5$ without charge density wave}

\author{Haochen Sun,$^{1,2,*}$ Jing Wang,$^{1,*}$ Deng Hu,$^{3,*}$ Zhiwei Wang,$^{3,\dagger}$ Shiyan Li$^{1,2,4,5,\ddagger}$}

\affiliation
{$^1$State Key Laboratory of Surface Physics, and Department of Physics, Fudan University, Shanghai 200438, China\\
 $^2$Shanghai Research Center for Quantum Sciences, Shanghai 201315, China\\
 $^3$Center for Quantum Physics, Key Laboratory of Advanced Optoelectronic, Quantum Architecture and Measurement (MOE), \\School of Physics, Beijing Institute of Technology, Beijing 100081, China\\
 $^4$Shanghai Branch, Hefei National Laboratory, Shanghai 201315, China\\
 $^5$Collaborative Innovation Center of Advanced Microstructures, Nanjing 210093, China
}

\date{\today}

\begin{abstract}

    The vanadium-based kagome superconductor CsV$_3$Sb$_5$ exhibits an intricate interplay between charge density wave (CDW) and superconductivity (SC). Substituting Ta for V will suppress the CDW order and enhance superconductivity. Here, we report a high-pressure study up to 48 GPa on Ta-doped Cs(V$_{0.86}$Ta$_{0.14}$)$_3$Sb$_5$ with enhanced $T_\mathrm{c}$ around 5.29 K and without CDW order. At low pressures, unlike the characteristic $``$M-shaped" temperature-pressure phase diagram of the pristine CsV$_3$Sb$_5$, the Ta-doped compound exhibits a distinct non-monotonic evolution. Initially, $T_\mathrm{c}$ drops slightly from 5.29 K at ambient pressure to 4.90 K at 0.33 GPa, before increasing to a maximum of 5.70 K at 0.61 GPa. Upon further compression, $T_\mathrm{c}$ decreases continuously to 0.98 K at 12.3 GPa. At higher pressures, $T_\mathrm{c}$ rises again, reaching 4.46 K at 47.7 GPa. This evolution of $T_\mathrm{c}$ establishes two superconducting domes under pressure, which are similar to those in CsV$_3$Sb$_5$. Our results demonstrate that pressure-induced double-dome superconductivity is intrinsic to these V-based kagome superconductors, even in the absence of long-range CDW order.

\end{abstract}

\maketitle
\section{Introduction}
%intro of AVS;
The vanadium-based kagome metals $A$V$_3$Sb$_5$ ($A$ = Cs, Rb, and K) have emerged as a prototypical platform to explore correlated physics, characterized by a unique electronic structure of flat bands, Dirac points, and van Hove singularities (VHSs) \cite{PRM2019ORTIZ, PRL2020ORTIZ, PRL2021TAN, NATURE2022YIN, NRM2024ORTIZ}.
%found-PRM 2019, 3, 094407; PRL 125, 247002 (2020);elec-structure-PRL 127, 046401 (2021).;review-Nature 2022, 612, 647.;Nature Reviews Materials volume 9, pages420–432 (2024).  
Within this family, a variety of exotic quantum states emerge, including superconductivity (SC), unconventional charge density wave (CDW), and electronic nematicity \cite{PRL2020ORTIZ, NRM2024ORTIZ, PRM2021ORTIZ, CPL2021YIN, PRL2021TAN, NM2021JIANG, PRX2021LI, NC2021XIANG, NP2022NEUPERT}.
%Cs-PRL 125, 247002 (2020);Nature Reviews Materials volume 9, pages420–432 (2024);Rb-PRM. 5, 034801 (2021);K-CPL 38, 037403 (2021).elec-structure-PRL 127, 046401 (2021).K-CDW-NM. 20, 1353 (2021).;Cs-RbPhysRevX.11.031050;NEMATICITY(2021)12:6727;Nature Physics (Neupert);
In CsV$_3$Sb$_5$, specifically, the material first transitions into a three-dimensional (3D) triple-$Q$ CDW state at $T_{\text{CDW}} \approx 94$ K \cite{PRL2020ORTIZ, PRX2021LI, PRX2021LIANG}.
%PhysRevLett.125.247002(2020);Phys. Rev. X 11, 031050 (2021);PhysRevX.11.031026(2021)
Upon further cooling, it develops an electronic nematic order around $T_\text{nem}\approx35\ \text{K}$ \cite{NC2021XIANG, NATURE2022NIE},
%NC(2021)12:3645;Nature 604,59–64 (2022)
then finally enters the superconducting phase at $T_\text{c}\approx2.5$ K \cite{PRL2020ORTIZ}.
%PhysRevLett.125.247002(2020);

Hydrostatic pressure serves as an effective tuning parameter in the study of $A$V$_3$Sb$_5$. In CsV$_3$Sb$_5$, $T_\text{CDW}$ is monotonically suppressed with increasing pressure up to 2 GPa, while $T_\text{c}$ exhibits a characteristic $``$M-shaped" evolution. Specifically, $T_\text{c}$ displays two maxima at critical pressures that coincide with the modification and eventual collapse of the CDW order \cite{PRL2021CHEN, NC2021YU}. 
%PRL. 126, 247001;NC(2021)12:3645
The non-monotonic behavior of $T_\text{c}$ suggests a strong interplay between the CDW and SC states. High-pressure nuclear magnetic resonance (NMR) and transport measurements reveal that the CDW changes from a triple-$Q$ to a stripe-like order at the first critical pressure $P_\text{c1}\approx 0.58$ GPa \cite{NATURE2022ZHENG}. This stripe-like CDW is eventually suppressed at $P_\text{c2}\approx2.0$ GPa, where $T_\mathrm{c}$ reaches its maximum of $~\sim$8 K \cite{NATURE2022ZHENG}.
%Nature  611, 682 (2022) 
The suppression of SC between $P_\text{c1}$ and $P_\text{c2}$ is thus closely associated with the emergence of the stripe-like CDW, pointing to a fierce competition between these two electronic phases.
At higher pressures, superconductivity reemerges above 16.5 GPa \cite{ARXIV2021ZHAO, PRB2021ZHANG, CPL2021CHEN}, 
%arXiv:2102.08356.;PhysRevB.103.224513;CPL. 38 057402
forming a double-dome phase diagram that persists up to 150 GPa \cite{PRL2022YU}. 
%PhysRevLett.128.077001
The first superconducting dome (below 10 GPa) is widely attributed to the competition between CDW and SC \cite{PRL2021CHEN, NC2021YU, NATURE2022ZHENG},
%PRL. 126, 247001;NC(2021)12:3645;Nature  611, 682 (2022).
 whereas the second dome is likely driven by a pressure-induced dimensional crossover and enhanced electron-phonon coupling \cite{PRL2022YU, CPL2021CHEN}.
    %%PhysRevLett.128.077001;CPL. 38 057402
Furthermore, similar double-dome superconducting phase diagrams are also observed in the sister compounds KV$_3$Sb$_5$ and RbV$_3$Sb$_5$ \cite{PRB2021DU, PRR2021WANG, PRB2022ZHU},
%PRB,103, L220504 (2021);Phys. Rev. Res. 3, 043018 (2021);PhysRevB.105.094507
 suggesting that this double-dome feature is a universal high-pressure behavior within this family.  

Complementary to hydrostatic pressure, chemical substitution serves as another powerful approach to tune the interplay between CDW and SC by introducing chemical pressure, charge carriers, and disorder. Substituting various elements at specific crystallographic sites allows for multiple modulations of the interplay between CDW and SC in $A$V$_3$Sb$_5$ \cite{NRM2024ORTIZ, SSAT2024ZHOU}. 
%Nature Reviews Materials volume 9, pages420–432 (2024);Supercond. Sci. Technol. 37 (2024) 103001
For instance, Sn substitution at the Sb site and Ti substitution at the V site in CsV$_3$Sb$_5$ both effectively suppress the long-range CDW order and result in double-dome superconductivity in the phase diagram \cite{PRM2022OEY, SCIBULL2022YANG, PRM2023LIU}.
%PRM 6, L041801 (2022);Sci Bull 67 (2022) 2176–2185;PRM 7, 064801 (2023)
High-resolution angle-resolved photoemission spectroscopy (ARPES) measurements, combined with first-principles calculations, reveal that hole-doping via Ti substitution lowers the Fermi level, gradually shifting the VHS above $E_F$ \cite{PRM2023LIU}. This quenches the CDW instability, explaining the rapid suppression of the CDW state \cite{PRM2023LIU}.
%PRM 7, 064801 (2023);
Furthermore, in both Sn- and Ti-doped CsV$_3$Sb$_5$, a distinct short-range charge stripe order emerges upon the collapse of the long-range CDW, which is closely associated with the suppression of SC in the intermediate doping regime \cite{PRX2025HUAI, SCIBULL2022YANG}.
%PRX 15, 041039 (2025);Sci Bull 67 (2022) 2176
 
Beyond carrier doping, extensive research has focused on the isovalent substitution of Ta for V in CsV$_3$Sb$_5$. Increasing the Ta concentration monotonically suppresses $T_\text{CDW}$ while simultaneously enhances superconductivity \cite{PRB2022LIU, PRB2022LI, SCIREP2024LIU}.
%PRB 106, L140501 (2022);PRB 106, 214529 (2022);SR(2024) 14:9580
 %X-Ray diffraction (XRD) experiments revealed 
ARPES measurements reveal that the VHS around the $M$ point coincides with the Fermi level in CsV$_{2.6}$Ta$_{0.4}$Sb$_5$, demonstrating a direct connection between the significantly enhanced SC and the alignment of VHS to the Fermi level \cite{NC2023LUO}.
%NC | (2023)14:3819
Furthermore, ultralow-temperature heat transport and scanning tunneling microscopy (STM) measurements show a fully gapped state with no residual density of states (DOS) at low temperature, indicating nodeless $s$-wave SC in Cs(V$_{0.86}$Ta$_{0.14}$)$_3$Sb$_5$ without CDW, contrasting with the possible unconventional pairing in its parent compound CsV$_3$Sb$_5$ with long-range CDW \cite{CPL2024ZHAO}.
%CPL 41, 127303 (2024);
While the high-pressure evolution of superconductivity driven by the interplay between CDW and SC has been extensively studied in the parent $A$V$_3$Sb$_5$, it will be interesting to examine whether the double-dome superconducting phase diagram is still there after the long-range CDW order is completely suppressed in doped $A$V$_3$Sb$_5$.  

In this paper, we measure the resistance of Cs(V$_{0.86}$Ta$_{0.14}$)$_3$Sb$_5$ without CDW order under pressure up to 48 GPa. A double-dome superconductivity is clearly revealed in the temperature-pressure phase diagram, which is similar to that observed in CsV$_3$Sb$_5$. We discuss the underlying physics of the two superconducting domes, proposing that the second dome is likely also associated with the dimensional crossover induced by pressure. Additionally, we suggest that the non-monotonic evolution of superconductivity at low pressure is linked to pressure-driven modifications of the electronic structure near the Fermi level.

\section{METHODS}
%\section{methods}
Single crystals of Cs(V$_{0.86}$Ta$_{0.14}$)$_3$Sb$_5$ were synthesized via a self-flux method using Cs bulk (Alfa Aesar 99.8\%), V pieces (Aladdin, 99.97\%), Ta powder (Alfa Aesar, 99.99\%), and Sb shot (Alfa Aesar, 99.9999\%) \cite{SCIREP2024LIU}. The Ta doping level was determined by energy dispersive X-ray spectroscopy (EDX) as in Ref. \cite{SCIREP2024LIU}. Direct-current (dc) magnetization measurements were performed down to 2.4 K using a Magnetic Property Measurement System (MPMS, Quantum Design). In-plane resistivity measurement of Cs(V$_{0.86}$Ta$_{0.14}$)$_3$Sb$_5$ single crystal was measured in a Physical Property Measurement System (PPMS, Quantum Design) with conventional four-probe method. 

High-pressure experiments were conducted using a BeCu-alloy diamond anvil cell (DAC) with culet sizes of 800 and 300 $\mu$m, respectively. Silicone oil and KBr were employed as the liquid and solid pressure-transmitting media (PTM), respectively. The pressure inside the DAC was calibrated via the ruby fluorescence method at room temperature before and after each experiment. High-pressure electrical resistance of Cs(V$_{0.86}$Ta$_{0.14}$)$_3$Sb$_5$ was measured using a standard van der Pauw configuration in a PPMS and a $^3$He cryostat (Oxford Instruments).
\section{RESULTS AND DISCUSSION}
%\section{RESULTS}
%--------------------------------------------Fig1----------------------------------------------------------------------------------------------
\begin{figure}
  \includegraphics[clip,width=7cm]{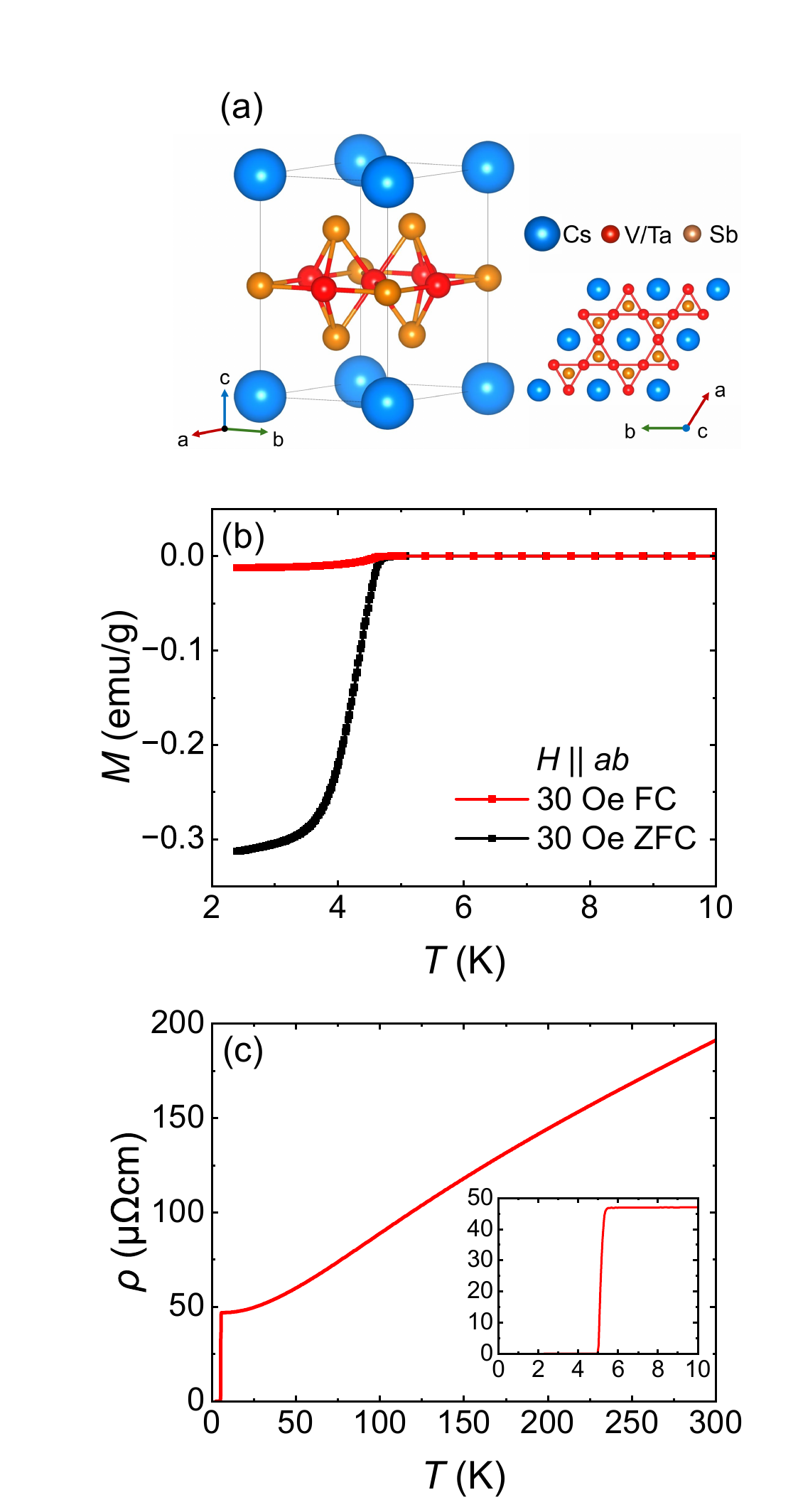}
  \caption{(a) Crystal structure of Cs(V$_{0.86}$Ta$_{0.14}$)$_3$Sb$_5$. The V/Ta atoms form layers with an ideal kagome lattice.  (b) DC magnetization of Cs(V$_{0.86}$Ta$_{0.14}$)$_3$Sb$_5$ single crystal at low temperatures, with zero-field and field cooling modes. (c) In-plane resistivity of Cs(V$_{0.86}$Ta$_{0.14}$)$_3$Sb$_5$ single crystal at ambient pressure. Inset shows the low-temperature superconducting transition.}
  \label{fig.1}
\end{figure}
Cs(V$_{0.86}$Ta$_{0.14}$)$_3$Sb$_5$ crystallizes in a layered hexagonal structure ($P6/mmm$), as shown in Fig.\,1(a). The V/Ta atoms form layers with an ideal kagome lattice. Figure 1(b) displays the dc magnetization measured from 2.4 to 10 K under zero-field-cooled (ZFC) and field-cooled (FC) modes. A distinct diamagnetic transition is observed at $T_\text{c}^{\text{onset}}\approx4.7$ K, confirming the enhancement of SC relative to CsV$_3$Sb$_5$. The temperature dependence of in-plane resistivity, as plotted in Fig. 1(c), shows typical metallic behavior with a residual resistivity ratio $RRR = 4.22$. This value is significantly lower than that of high-quality CsV$_3$Sb$_5$ single crystal \cite{PRB2022CHEN}, 
%Phys. Rev. B 105, L201109 (2022)%
indicating the significant impact on transport properties from chemical disorders introduced by Ta substitution. Moreover, the absence of any resistivity anomaly associated with the CDW transition in the $\rho(T)$ curve is consistent with previous reports \cite{SCIREP2024LIU,CPL2024ZHAO}, confirming the complete suppression of the long-range CDW order at this doping level. The low-temperature resistivity in the inset of Fig. 1(c) shows a clear superconducting transition with zero resistivity at $T_\mathrm{c}^{\text{zero}}$ = 4.92 K.

\begin{figure}
\centering
  \includegraphics[clip,width=5.675cm]{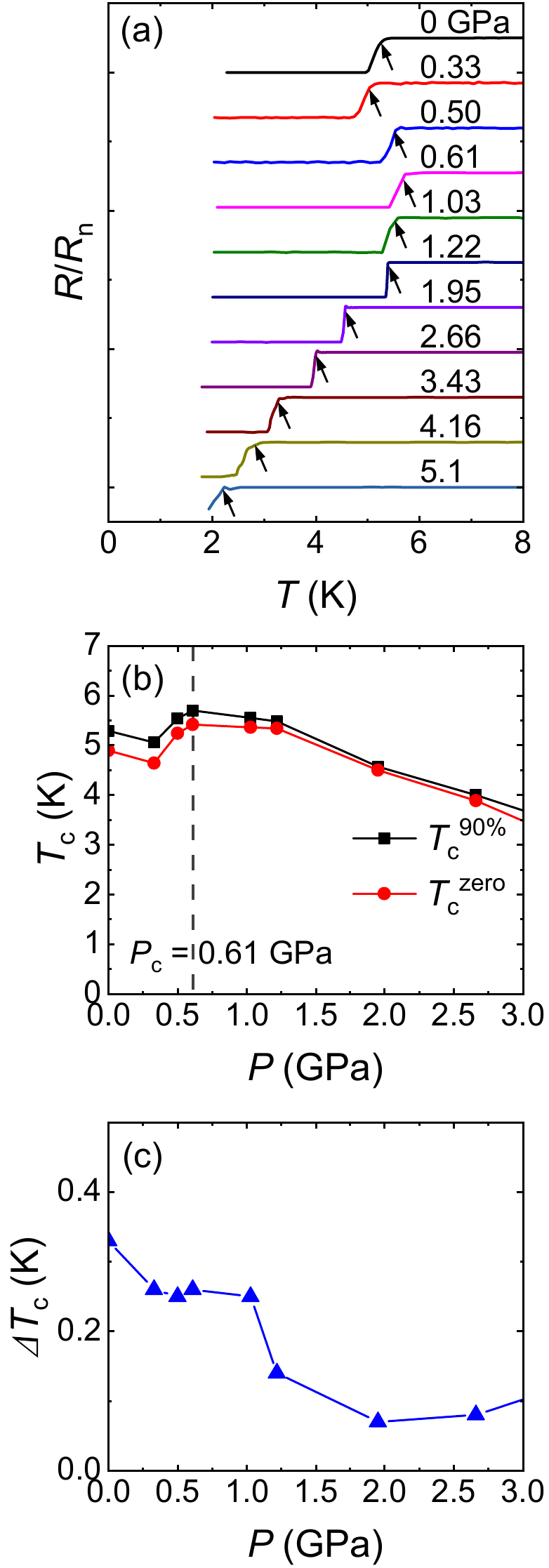}
  \label{fig.2}
  \caption{(a) The temperature dependence of the normalized resistance for Cs(V$_{0.86}$Ta$_{0.14}$)$_3$Sb$_5$ under various pressures up to 5.1 GPa. $R_{\mathrm{n}}$ represents the normal-state resistance of the sample right above $T_c$. The arrows indicate the superconducting transition temperature $T_\mathrm{c}^{90\%}$. (b) Pressure dependence of the superconducting transition temperatures $T_\mathrm{c}^{90\%}$ and $T_\mathrm{c}^{\text{zero}}$ in the low-pressure regime below 3 GPa. (c) The superconducting transition width $\Delta T_\text{c}$ at various pressures, defined as $T_\mathrm{c}^{90\%}$ - $T_\mathrm{c}^{10\%}$. 
}
\end{figure}

\begin{figure}
\centering
  \includegraphics[clip,width=6cm]{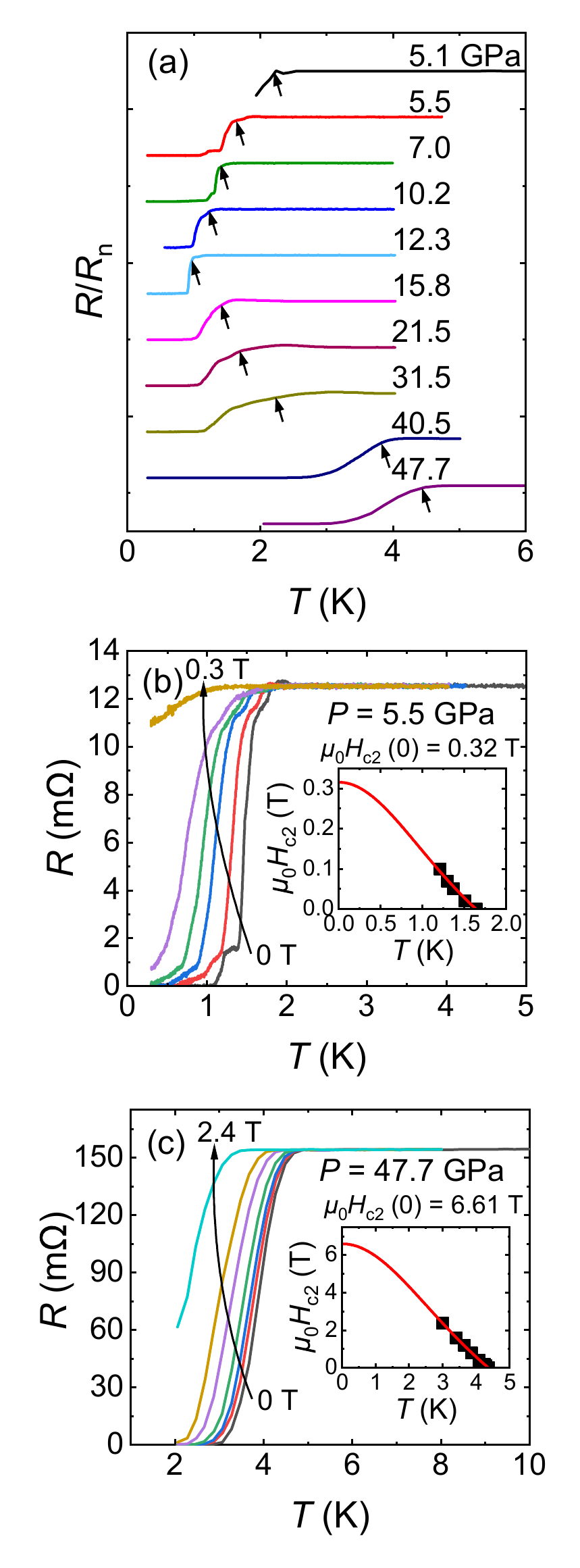}
  \caption{(a) Temperature dependence of the normalized resistance for Cs(V$_{0.86}$Ta$_{0.14}$)$_3$Sb$_5$ under pressures from 5.1 to 47.7 GPa. (b) and (c) Temperature dependence of resistance for Cs(V$_{0.86}$Ta$_{0.14}$)$_3$Sb$_5$ under different magnetic fields at 5.5 GPa and 47.7 GPa, respectively. The magnetic field gradually suppresses the superconducting transition. Insets: Temperature dependence of the upper critical field $\mu_{\mathrm{0}} H_{\mathrm{c2}}$ at 5.5 GPa and 47.7 GPa, respectively. The red lines are the fitting of the data to the Ginzburg-Landau formula.}
\end{figure}

Figure 2(a) displays the low-temperature normalized resistance of a Cs(V$_{0.86}$Ta$_{0.14}$)$_3$Sb$_5$ sample from ambient pressure (AP) to 5.1 GPa, measured with 800 $\mu$m culet and silicone oil as the PTM. The superconducting transition temperature, defined at 90\% of the normal-state resistance ($T_\mathrm{c}^{90\%}$), initially drops slightly from 5.29 K at AP to 4.9 K at 0.33 GPa, before rising to a maximum of 5.7 K at 0.61 GPa. Upon further compression, $T_\mathrm{c}$ decreases continuously with increasing pressure up to 5.1 GPa. This distinct non-monotonic evolution of superconductivity at low pressures below 3 GPa, shown in Fig. 2(b), contrasts with the $``$M-shaped" phase diagram in parent CsV$_3$Sb$_5$. Meanwhile, it also differs from the simple dome-shaped superconductivity observed in Ti-doped CsV$_{2.85}$Ti$_{0.15}$Sb$_5$ at low pressures \cite{PRB2023HOU}.

To map out the complete phase diagram, we extend the high-pressure resistance measurements up to 48 GPa, with 300 $\mu$m culet and KBr as the PTM. Figure 3(a) displays the low-temperature normalized resistance of the Cs(V$_{0.86}$Ta$_{0.14}$)$_3$Sb$_5$ sample from 5.1 to 47.7 GPa. The data under pressure of 5.1 GPa are from Fig. 2(a). One can see that the $T_\mathrm{c}$ continuously decreases to 0.98 K at 12.3 GPa upon further compression. Beyond 12.3 GPa, $T_\mathrm{c}$ starts to rise again, reaching 4.46 K at the highest pressure of 47.7 GPa. The notably broadened superconducting transition width at higher pressures can be attributed to the pressure gradient within the diamond anvil cell.

To confirm the superconducting nature of the resistance drop in Cs(V$_{0.86}$Ta$_{0.14}$)$_3$Sb$_5$ under high pressures, we perform resistance measurements under various magnetic fields for 5.5 and 47.7 GPa, as depicted in Fig. 3(b) and 3(c). The resistance drop is gradually shifted to lower temperatures with increasing field, confirming its superconducting origin. The insets of Fig. 3(b) and 3(c) show the temperature dependence of the upper critical field $\mu_0 H_{\mathrm{c2}}$. The data $\mu_0 H_{\mathrm{c2}}(T)$ can be well fitted by the empirical Ginzburg-Landau formula $\mu_0H_{\text{c}2}(T) = \mu_0H_{\text{c}2}(0)[1 - (T/T_{\text{c}})^2]/[1 + (T/T_{\text{c}})^2]$. The extracted $\mu_0H_{\text{c}2}(0)$ are 0.32 T for 5.49 GPa and 6.61 T for 47.7 GPa. Both values are well below their respective Pauli paramagnetic limit $H_\text{p} = 1.84T_\text{c}$, which are calculated as $3.02\ \text{T}$ and $8.04\ \text{T}$, suggesting the absence of Pauli pair-breaking effect.

Based on the high-pressure resistance measurements above, we construct the temperature-pressure ($T$-$P$) phase diagram in Fig. 4. The phase diagram reveals two distinct superconducting domes for Cs(V$_{0.86}$Ta$_{0.14}$)$_3$Sb$_5$, very similar to that of CsV$_3$Sb$_5$ over a wide pressure range \cite{ARXIV2021ZHAO, PRB2021ZHANG, CPL2021CHEN}, even though there is no CDW order in Cs(V$_{0.86}$Ta$_{0.14}$)$_3$Sb$_5$. 

Given that all three parent compounds in the $A$V$_3$Sb$_5$ family exhibit such a double-dome SC under pressure, the $T$-$P$ phase diagram of Cs(V$_{0.86}$Ta$_{0.14}$)$_3$Sb$_5$ extends this ubiquitous feature to chemically doped variants in this kagome system. However, the underlying mechanisms driving the double-dome SC across the family is still under debate. In KV$_3$Sb$_5$ and RbV$_3$Sb$_5$, structural phase transitions occur near the minima of the $T_{\text{c}}(P)$ curves, which are believed to account for the formation of two superconducting domes in both materials \cite{PRB2022DU}. In contrast, the lattice symmetry of CsV$_3$Sb$_5$ remains stable up to 80 GPa, indicating that the emergence of its second superconducting dome cannot be attributed to an abrupt structural phase transition \cite{PRL2022YU, SCIPOST2022TSIRLIN}. Instead, a pressure-induced dimensional crossover has been observed in CsV$_3$Sb$_5$ \cite{PRL2022YU}. The formation of interlayer Sb-Sb bonding leads to a reconstruction of Sb-derived bands around 10 GPa, which enhances the density of states at the Fermi level \cite{PRL2022YU}. This mechanism has been proposed to drive the enhancement of $T_c$ in the second superconducting dome \cite{PRL2022YU}. Furthermore, a recent theoretical study on superconductivity driven by soft loop-current fluctuations in kagome metals suggests that the Sb-dominated Fermi surface sheet undergoes a Lifshitz transition with increasing pressure, driving a transition of SC from an $s^\pm$ state (low-pressure SC-I) to a chiral $d+id$ state (high-pressure SC-II) \cite{NC2026SCHULTZ}. Putting together, we propose that the emergence of the second superconduting dome in Cs(V$_{0.86}$Ta$_{0.14}$)$_3$Sb$_5$ is similarly driven by a pressure-induced Fermi surface Lifshitz transition, as in the pristine CsV$_3$Sb$_5$.

\begin{figure}
\centering
  \includegraphics[clip,width=7cm]{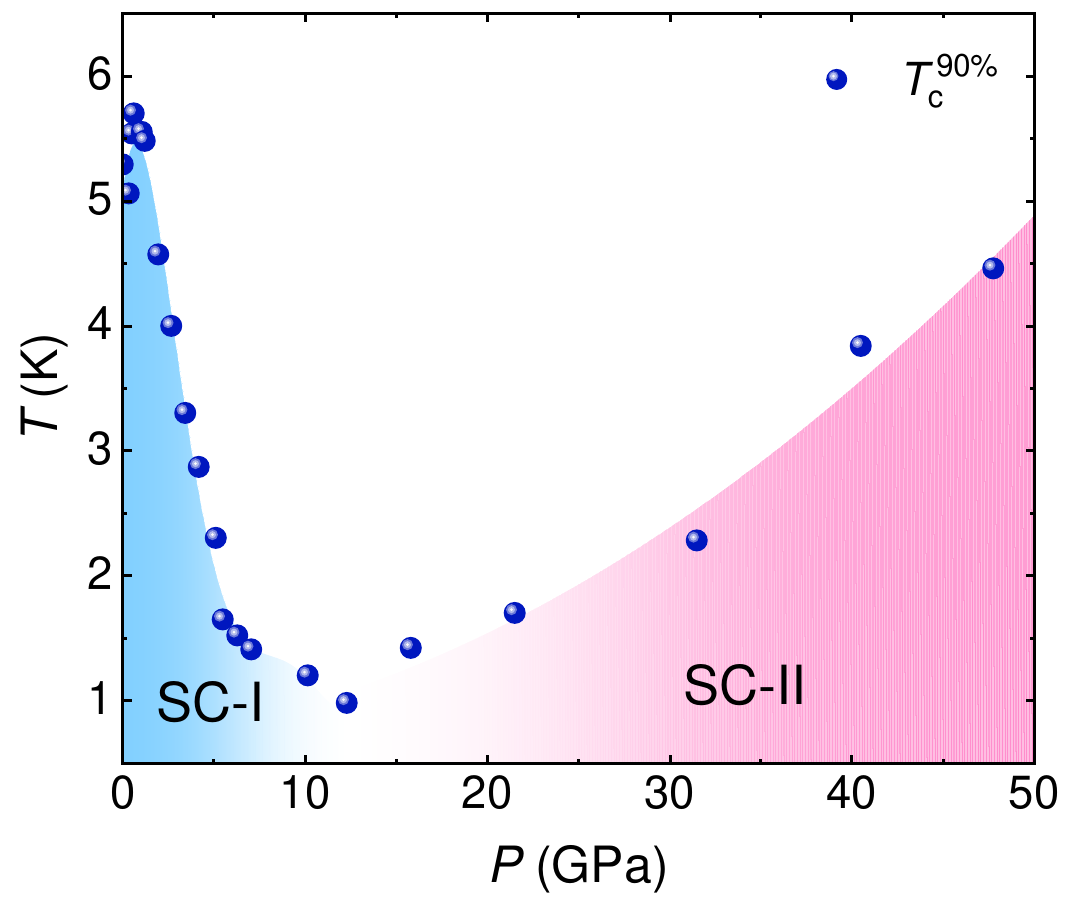}
  \caption{Temperature-pressure phase diagram of Cs(V$_{0.86}$Ta$_{0.14}$)$_3$Sb$_5$. Two superconducting domes are shown clearly, with SC-I phase at lower pressures and SC-II phase at higher pressures.}
\end{figure}

At low pressures, the non-monotonic evolution of SC in Cs(V$_{0.86}$Ta$_{0.14}$)$_3$Sb$_5$ is very intriguing. Unlike the parent CsV$_3$Sb$_5$, which exhibits a unique $``$M-shaped" superconducting dome with two maxima below 3 GPa, the $T_c$ of Cs(V$_{0.86}$Ta$_{0.14}$)$_3$Sb$_5$  initially decreases with pressure before rising to a maximum of 5.7 K at 0.61 GPa (Fig. 2). In CsV$_3$Sb$_5$, transport and NMR studies have indicated that the suppression of $T_c$ and the broadening of the transition around 0.58 GPa are correlated with a transition from a triple-$Q$ to a stripe-like CDW, highlighting a fierce competition between the SC and the long-range CDW \cite{PRL2021CHEN, NATURE2022ZHENG}. Consequently, the absence of long-range CDW order in Cs(V$_{0.86}$Ta$_{0.14}$)$_3$Sb$_5$ naturally precludes this specific $``$M-shaped" dome in CsV$_3$Sb$_5$.

However, this non-monotonic low-pressure behavior in the Ta-doped sample also distinctly differs from that of the Ti-doped CsV$_{2.85}$Ti$_{0.15}$Sb$_5$ sample. In CsV$_{2.85}$Ti$_{0.15}$Sb$_5$, where the long-range CDW is also suppressed by doping, the $``$M-shaped" dome is replaced by a simple smooth superconducting dome \cite{PRB2023HOU}, which is attributed to the pressure-induced suppression of short-range CDW correlations \cite{SCIBULL2022YANG, PRB2023HOU}. Furthermore, recent studies, which reveal short-range CDW fluctuations in Ti- and Sn-doped CsV$_3$Sb$_5$, suggest that such CDW fluctuations represent an intrinsic state that survives deep into the disordered system \cite{NPJ2023LINUS, PRL2025HUANG, PRM2026KIT}. Based on these observations, we postulate that the non-monotonic $T_c$ evolution within the first dome of Cs(V$_{0.86}$Ta$_{0.14}$)$_3$Sb$_5$ is likely associated with the interplay between SC and elusive short-range CDW fluctuations. 

Beyond the scenario of competing orders, recent studies have demonstrated that the superconducting $T_c$ is intimately related to the energy position of the VHS in various doped CsV$_3$Sb$_5$ samples lacking long-range CDW order \cite{NC2023LUO}. For instance, the significant enhancement of superconductivity in CsV$_{2.6}$Ta$_{0.4}$Sb$_5$ is attributed to the alignment of VHS with the Fermi level at the $M$ point \cite{NC2023LUO}. Therefore, it is possible that the non-monotonic evolution of $T_c$ observed in the first dome may also correlate to a pressure-induced shift of the VHS relative to the Fermi level. The exact microscopic mechanism governing this low-pressure behavior still needs further investigation.

\section{Summary}

In summary, we perform high-pressure resistance measurements on Ta-doped Cs(V$_{0.86}$Ta$_{0.14}$)$_3$Sb$_5$ to check whether the double-dome superconductivity still exists when there is no CDW order. Two superconducting domes under pressure are clearly revealed in the $T-P$ phase diagram, reinforcing the notion that double-dome superconductivity is an intrinsic and universal feature of the $A$V$_3$Sb$_5$ family, even in the complete absence of long-range CDW. Notably, we observe a non-monotonic evolution of $T_c$ at low pressures, which contrasts with the $``$M-shaped" dome in the parent CsV$_3$Sb$_5$. While the high-pressure dome may come from a pressure-induced Fermi surface Lifshitz transition, the origin of the low-pressure non-monotonic $T_c$ evolution is still unclear. Our findings provide important insights into the pressure-tuned phase diagram of doped kagome metals, shedding new light on the interplay between charge order and superconductivity in kagome metals.

\acknowledgments

This work was supported by the Natural Science Foundation of China (Grant No. 12674162), the Shanghai Municipal Science and Technology Major Project (Grant No. 2019SHZDZX01), Quantum Science and Technology-National Science and Technology Major Project (Grant No. 2024ZD0300104), the National Key Research and Development 
Program of China (Grant No. 2022YFA1403400), and the Beijing National Laboratory for Condensed Matter Physics (Grant No. 2023BNLCMPKF007).\\

\noindent $^*$ These authors contributed equally to this work.\\
\noindent $^\dagger$ Contact author: zhiweiwang$@$bit.edu.cn\\
\noindent $^\ddagger$ Contact author: shiyan$\_$li$@$fudan.edu.cn

\end{document}